\documentclass[intlimits,twoside,a4paper]{article}

\usepackage{amsmath,amssymb}
\usepackage{graphicx}

\usepackage{wrapfig}

\usepackage[T2A]{fontenc}
\usepackage[cp1251]{inputenc}
\usepackage{cmpj2}

\issue{2012}{15}{1}{13603}

\doinumber{10.5488/CMP.15.13603}

\title{Elastic and electronic properties of fluorite RuO$_2$ from first principle}

\author[Z.J. Yang \textsl{et al.}]{Z.J. Yang\refaddr{ad1}\thanks{E-mail: yzjscu@163.com,  Tel: 86 28 85405526, Fax: 86 28 85405515}\,, A.M. Guo\refaddr{ad2},
Y.D. Guo\refaddr{ad3}, J. Li\refaddr{ad4}, Z. Wang\refaddr{ad4}, Q. Liu\refaddr{ad5}, R.F. Linghu\refaddr{ad6}, X.D. Yang\refaddr{ad7}
}

\addresses{\addr{ad1}School of Science, Zhejiang University of
Technology,  Hangzhou 310023, China
\addr{ad2} Department of Physics, California State University,
Northridge,  California 91330--8268, USA
\addr{ad3}School of Physics, Neijiang Normal University,
 Neijiang 641112,  China
\addr{ad4}College of Material and Chemical Engineering, Hainan
Provincial Key Laboratory of Research on Utilization of Si-Zr-Ti
Resources, Hainan University,  Haikou 570228, China
\addr{ad5}School of Physics, Chongqing University of Technology,
 Chongqing 400050, China
\addr{ad6}School of Physics, Guizhou Normal University,  Guiyang 550001,
 China
\addr{ad7}Institute of Atomic and Molecular Physics, Sichuan
University,  Chengdu 610065, China
}

\date{Received May 2, 2011, in final form February 1, 2012}

\authorcopyright{Z.J. Yang, A.M.~Guo, Y.D.~Guo, J.~Li, Z.~Wang, Q.~Liu, R.F.~Linghu, X.D.~Yang, 2012}

\begin{document}

\maketitle

\begin{abstract}
The elastic, thermodynamic, and electronic properties of fluorite
RuO$_2$ under high pressure are investigated by plane-wave
pseudopotential density functional theory. The optimized lattice
parameters, elastic constants, bulk modulus, and shear modulus are
consistent with other theoretical values. The phase transition from
modified fluorite-type to fluorite is 88~GPa (by localized density
approximation, LDA) or 115.5~GPa (by generalized gradient
approximation, GGA). The Young's modulus and Lam\'{e}'s coefficients
are also studied under high pressure. The structure turned out to be stable
for the pressure up to 120~GPa by calculating elastic constants. In
addition, the thermodynamic properties, including the Debye
temperature, heat capacity, thermal expansion coefficient,
Gr\"{u}neisen parameter, and Poisson's ratio, are investigated. A
small band gap is found in the electronic structure of fluorite
RuO$_2$ and the bandwidth increases with the pressure. Also, the
present mechanical and electronic properties demonstrate that the
bonding nature is a combination of covalent, ionic, and metallic
contributions.

\pacs 63.20.dk, 71.20.-b, 62.20.D-, 65.40.-b

\end{abstract}

\section{Introduction}

Attractions to study RuO$_2$ are due to its fundamental properties
and potential superhard characteristics~\cite{1,2}. An extensive
search for new superhard materials has been undertaken during recent
years and a new class of hard materials has been suggested: the
transition-metal dioxides containing heavy elements. Typically, the
bulk modulus of modified fluorite (pyrite phase, \textit{Pa}\={3})
RuO$_2$ was found to be 399~GPa~\cite{3}, which is the highest value
except for diamond (442~GPa)~\cite{4}. RuO$_2$ has the rutile (\textit{P}4$_2$/\textit{mnm}) structure under usual conditions~\cite{3}, and
can be transformed to an orthorhombic (CaCl$_2$-type, \textit{Pnnm})
structure at about 6~GPa~\cite{5} or 11.8~GPa~\cite{6} and to a
pyrite structure at about 12~GPa~\cite{2,5}. Moreover, the theory
indicates the \textit{Pa}\={3} structure can be transformed to a
fluorite (\textit{Fm}\={3}\textit{m}) structure at about 89~GPa or 97~GPa
\cite{2}.

Recently, elastic properties focusing on \textit{Pa}\={3} phase of
RuO$_2$ have been investigated systematically~\cite{2,7}. Electronic
structures~\cite{8,9,10,11} and optical properties~\cite{12,13,14}
of rutile and orthorhombic~\cite{14} RuO$_2$ have been extensively
studied. A full-potential linear muffin-tin orbital calculation on
the electronic structure and bulk modulus of fluorite RuO$_2$ has
also been performed~\cite{15}. The hardness and elasticity in cubic
RuO$_2$ and Raman scattering of the rutile-to-CaCl$_2$ phase
transition have been probed experimentally~\cite{16}.

Previous investigations on fluorite RuO$_2$ are not complete and
some problems remain unresolved. Many properties, such as the
hardness, stabilization, elastic and thermodynamic properties \textit{etc} under high pressure, are still unknown. To reveal the superhard
characteristics appropriately, a detailed theoretical description of
the elastic and electronic properties is necessary.

\section{Theoretical approaches}

In this work, all the calculations have been performed with CASTEP
\cite{17,18}. In the electronic structure calculations, we have used
the non-local ultrasoft pseudopotential~\cite{19}, together with the
revised Perdew-Burke-Ernzerhof (RPBE) generalized gradient
approximation (GGA) exchange-correlation function~\cite{20}.
Considering the computational cost, a plane-wave basis set with an
energy cut-off of $600.0$~eV~\cite{2} has been applied, and the
$12\times12\times12$ Monkhorst-Pack mesh has been used for the
Brillouin-zone (BZ) \textit{k}-point sampling. Pseudo atomic
calculations have been performed for Ru ($4s^24p^64d^75s^1$)
and O ($2s^22p^4$), where the self-consistent convergence of the
total energy is at $5.0\times10^{-7}$~eV/atom.

\section{Results and discussion}

In the equilibrium geometry calculations of fluorite RuO$_2$, both
the GGA and the LDA methods have been used. The bulk modulus ($B_0$)
and its first-order pressure derivative ($B_0'$) by Murnaghan
\cite{21} equation of state are listed in table~\ref{tab:table1}.
\begin{table}[ht]
\caption{\label{tab:table1}The calculated lattice constants \textit{a}
({\AA}), phase transition pressure $P_t$ (GPa) and elastic constants
$c_{11}$, $c_{44}$, $c_{12}$ (GPa) by LDA and GGA methods at 0~GPa
and 0 K. The bulk modulus $B$ and shear modulus $G$ are calculated
by the elastic constants. The bulk modulus $B_0$ (GPa) and its
first-order pressure derivatives $B_0'$ are fitted by the Murnaghan
equation of state.}
\vspace{2ex}
\begin{center}
{\small
\begin{tabular}{|c||ccccccccc|}
\hline
&$a$ &$B_0$ &$B_0'$ & $P_t$
&$c_{11}$ &$c_{44}$ &$c_{12}$ &$B$ & $G$ \\
\hline \hline LDA (present)& 4.7349 & 353 & 4.13 & 88
& 680.26 & 209.87 & 175.36 & 343.66 & 225.98 \\
GGA (present)& 4.8584 & 287 & 3.77 & 115.5
& 569.09 & 170.20 & 117.35 & 267.93 & 198.16 \\
(LDA+GGA)/2& 4.7967 & 320 & 3.95 & 102
& 624.68 & 190.04 & 146.36 & 305.79 & 212.07 \\
\cite{1,22}&  & 345 & & 65 &&&&&\\
\cite{2}& 4.743 & 384,351 & 3.5,4.2 & 89 &&&&&\\
\cite{2}& 4.842 & 336,297 & 3.5,4.1 & 97 & 435 & 152
& 227 & & 133 \\
\cite{15}& & 343 &&&&&&&\\
\cite{23}& 4.842 & 328 & 4.2 & & 410.6 & 62 & 286.6 & & 62 \\
\hline
\end{tabular}}
\end{center}
\hspace{0.6cm}
{\footnotesize The LDA/CAPZ (embedded in CASTEP) calculations are performed using
the same parameter input with GGA.}
\end{table}

\begin{figure}[ht]
\centerline{\includegraphics[width=8cm]{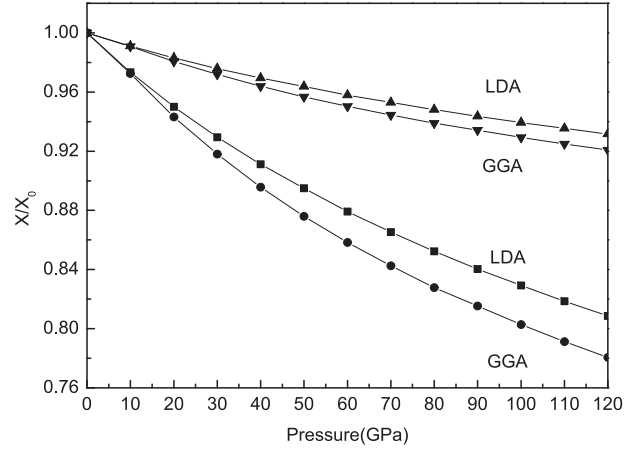}}
\caption{Variation of the lattice volume (denoted by square and
circle symbols) and the Ru-O bond length (denoted by triangle
symbols) with pressure. They are normalized by $X/X_0$, where $X$
and $X_0$ are the lattice volume or Ru-O bond length at any given
pressure and zero pressure at zero temperature. \label{fig:one}}
\end{figure}

Figure~\ref{fig:one} suggests that a significant stiffer
compressibility of 82.92\% is obtained using LDA at 100~GPa as
compared to GGA (80.27\%). The volume compressibility (about 80\%)
is nearly the same as the ultrastiff cubic TiO$_2$ at the same
pressure~\cite{24}. The bulk modulus of the fluorite RuO$_2$ is 353
GPa (LDA) or 287~GPa (GGA), which is slightly smaller than that of
the TiO$_2$ ($282~\text{GPa}\div395~\text{GPa}$). As the pressure is increased to
100~GPa, the approximation value of the normalized volume of
diamond, \textit{c}-BN, OsO$_2$, and OsC, is 85\%, 83\%, 80\%, and
84\%, respectively~\cite{25,26,27}, which is slightly larger than
the current calculations. As a comparison, figure~\ref{fig:one} shows
the Ru-O bond length contraction with increasing the pressure.
Similarly, a smaller Ru-O bond length contraction is obtained using
LDA. In a word, the current Ru-O bond length contraction, using
either LDA or GGA, is much smaller than that in volume contraction.

The calculated elastic constants of fluorite RuO$_2$ under different
pressures are listed in table~\ref{tab:table2}. According to the
generalized elastic stability criteria~\cite{28} ($c_{11}-c_{12}>0$,
$c_{11}+2c_{12}>0$, $c_{44}>0$) for cubic crystals, we can
demonstrate that fluorite structure is elastic stable under 120~GPa.
The larger $c_{11}$ and smaller $c_{12}$ indicate the
inter-atomic bonding along the \textit{c}-axis stronger than that along the \textit{a}-axis, consistent with the case of the larger bulk modulus \textit{B}
and the smaller shear modulus \textit{G} over a wide pressure range.
Compared to $c_{12}$ and $c_{44}$, $c_{11}$ varies largely by
changing the pressure, meaning that it is more difficult to obtain
the same strain from the longitudinal direction than from the
transverse direction.
\begin{table}[!b]
\caption{\label{tab:table2}The calculated (by GGA method) elastic
constants $c_{11}$, $c_{44}$, $c_{12}$ (GPa), heat capacity $C_V$
(\mbox{J$\cdot$ mol$^{-1}$$\cdot$K$^{-1}$}), Debye temperature
$\Theta$~(K), Gr\"{u}neisen parameter $\gamma$, thermal
expansion coefficient $\alpha$ (10$^{-5}$~K$^{-1}$) and
Poisson's ratio $\sigma$ over a wide pressure range at zero
temperature.}
\vspace{2ex}
\begin{center}
{\small
\begin{tabular}{|c||ccccccccc|}
\hline
$P$ &$V/V_0$ & $c_{11}$ &$c_{44}$ & $c_{12}$
&$C_V$ &$\Theta$ &$\gamma$ &$\alpha$ & $\sigma$ \\
\hline \hline 0& 1 & 569.0981 & 170.2064 & 117.3517
& 47.1572 & 872.0784 & 0.7086 & 4.3589 & 0.2124 \\
10& 0.9725 & 618.5525 & 182.6494 & 142.5632
& 42.9681 & 905.9000 & 0.6871 & 3.5224 & 0.2247 \\
20& 0.9432 & 672.0389 & 203.7597 & 177.3234
& 37.9710 & 948.9614 & 0.6643 & 2.7310 & 0.2351 \\
30& 0.9719 & 724.0007 & 214.6731 & 213.0271
& 34.9787 & 976.2347 & 0.6447 & 2.2390 & 0.2498 \\
40& 0.8957 & 780.0562 & 227.5737 & 254.8032
& 32.0164 & 1004.6197 & 0.6276 & 1.8230 & 0.2638 \\
50& 0.8759 & 832.5314 & 239.1418 & 287.9699
& 29.2934 & 1031.9559 & 0.6124 & 1.5240 & 0.2724 \\
60& 0.8583 & 882.1365 & 250.6513 & 326.8776
& 27.0493 & 1055.4048 & 0.5989 & 1.2880 & 0.2820 \\
70& 0.8426 & 928.6125 & 258.9245 & 359.1524
& 25.1941 & 1075.4502 & 0.5870 & 1.1170 & 0.2894 \\
80& 0.8278 & 985.5626 & 273.7571 & 401.9691
& 22.7703 & 1103.0287 & 0.5757 & 0.9278 & 0.2965 \\
90& 0.8153 & 1024.6117 & 279.8891 & 438.2222
& 21.7231 & 1115.3097 & 0.5663 & 0.8321 & 0.3043 \\
100& 0.8027 & 1066.3426 & 285.4165 & 469.7282
& 20.5262 & 1129.5911 & 0.5568 & 0.7442 & 0.3102 \\
110& 0.7913 & 1111.9641 & 294.8567 & 508.1361
& 19.1198 & 1146.9061 & 0.5482 & 0.6525 & 0.3159 \\
120& 0.7806 & 1156.7986 & 294.2600 & 540.1705
& 18.4615 & 1154.9119 & 0.5401 & 0.5986 & 0.3227 \\
\hline\end{tabular}}
\end{center}
\end{table}

\begin{figure}[ht]
\includegraphics[width=0.48\textwidth]{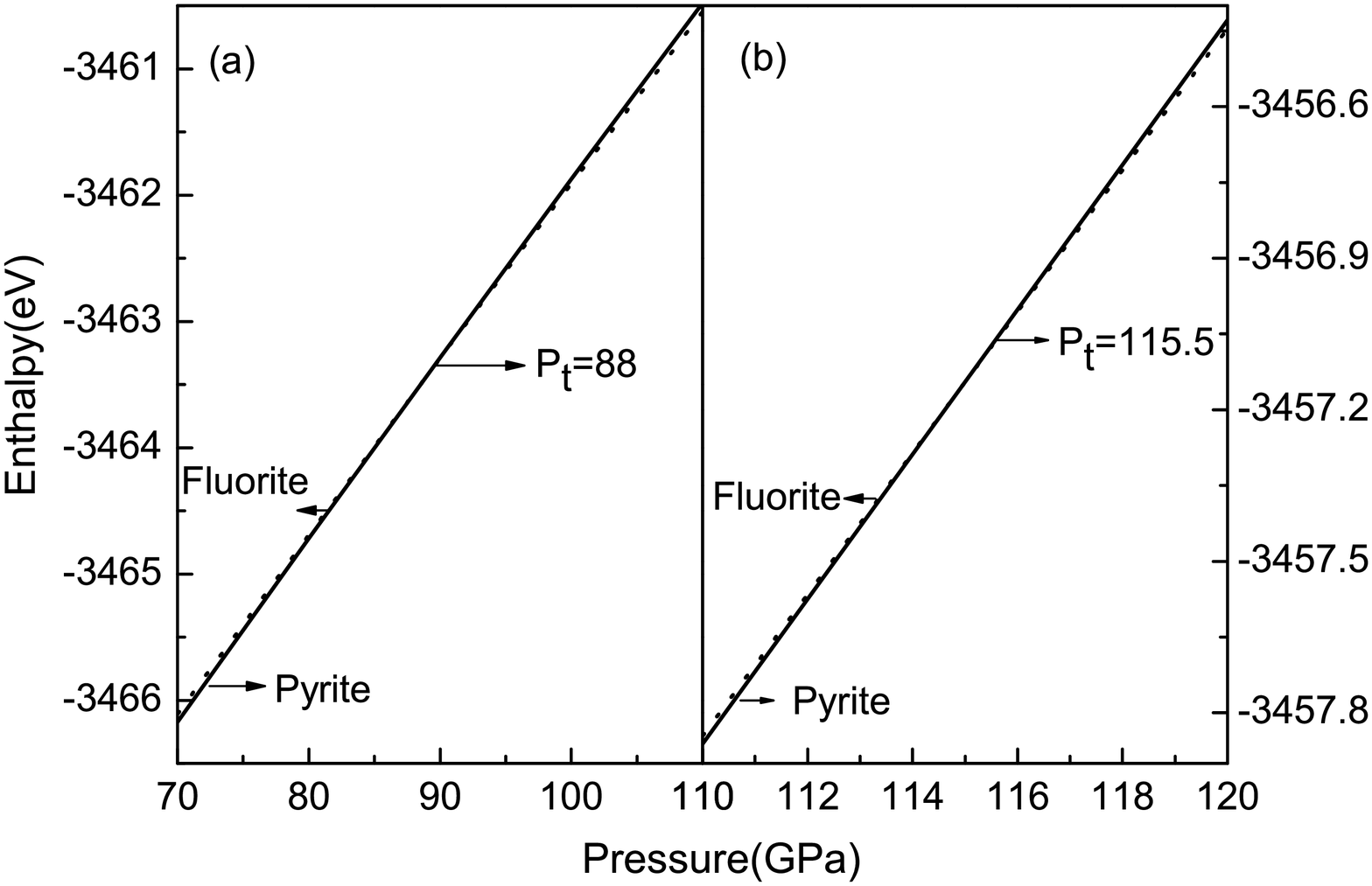}%
\hfill%
\includegraphics[width=0.41\textwidth]{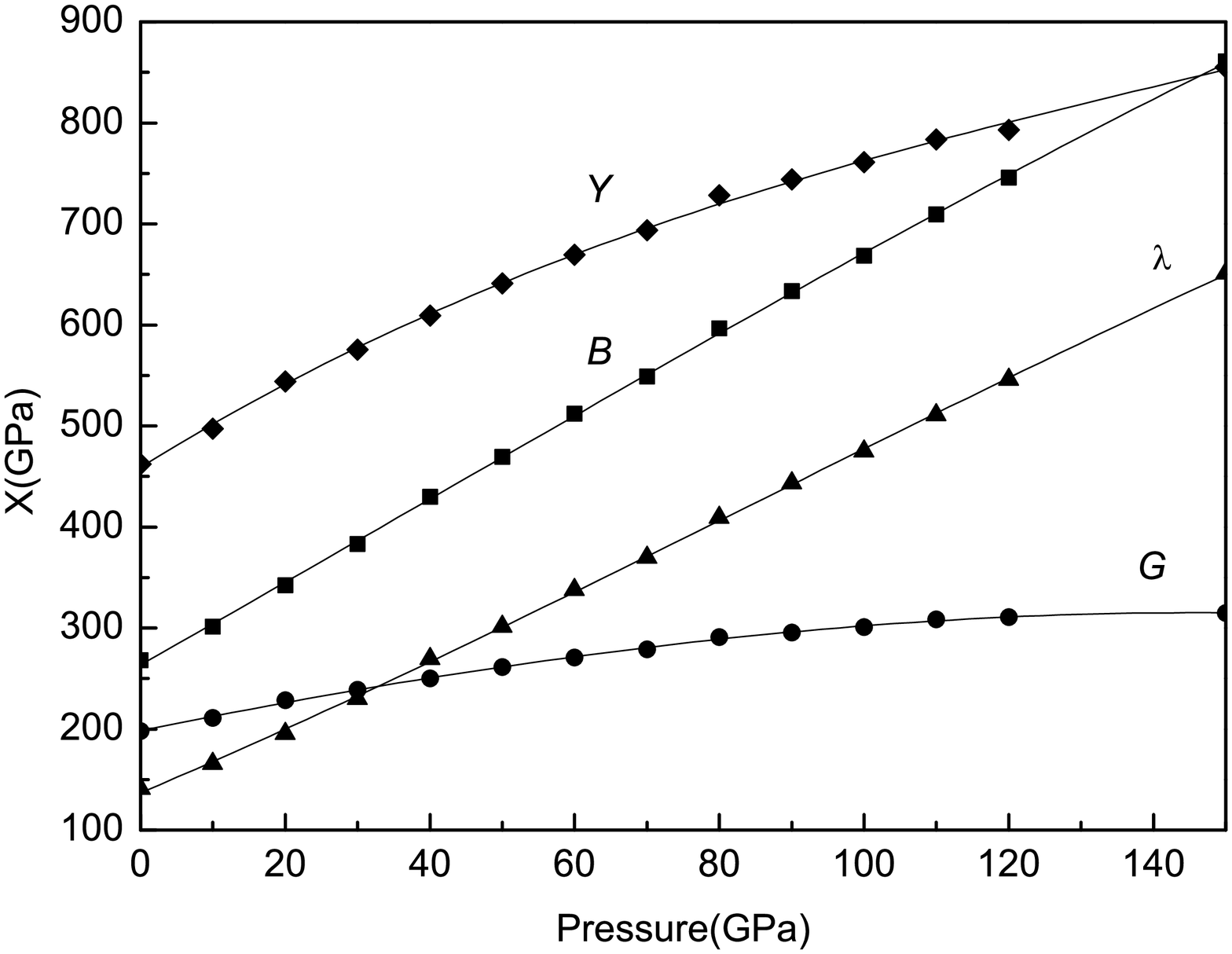}%
\hspace{0.8cm}
\\%
\parbox[t]{0.48\textwidth}{%
\caption{Enthalpy as a function of pressures for the \textit{Pa}\={3}
and \textit{Fm}\={3}\textit{m} phases of RuO$_2$, (a) is the LDA results
and (b) is GGA ones. \label{fig:two}}%
}%
\hfill%
\parbox[t]{0.48\textwidth}{%
\caption{Pressure dependences of mechanical quantities (by GGA
method) under different pressures, $X$ represents Bulk modulus $B$,
Shear modulus $G$, Young's modulus $Y$, Lam\'{e}'s coefficients
$\lambda$. \label{fig:three}}%
}%
\end{figure}

An estimate of the zero-temperature transition pressure between the
\textit{Pa}\={3} and \textit{Fm}\={3}\textit{m} structures may be obtained
from the usual condition of equal enthalpies, i.e., the pressure
\textit{P}, at which enthalpy $H=E+PV$ of both phases is the same. Our
calculated \textit{Pa}\={3}$\rightarrow$\textit{Fm}\={3}\textit{m} phase
transition pressure is 115.5~GPa by GGA and 88~GPa by LDA, as shown in figure~\ref{fig:two}, in
accordance with the theoretical values of 89~GPa~\cite{2} and 97~GPa
\cite{2} but is greater than the predicted 65~GPa~\cite{1,22}.

Using the calculated elastic constants at 0 K and 0~GPa, we obtain
the bulk moduli \textit{B} of fluorite RuO$_2$, with the values of 287
GPa (GGA) and 353~GPa (LDA), respectively, which is smaller than
that of diamond (442~GPa~\cite{27}), although both of them have
comparable compressibility at 100~GPa. The shear constant $c_{44}$
is 170.20~GPa (GGA), which is consistent with previous theoretical
calculations 152~GPa~\cite{2}, 140~GPa~\cite{16}, 147~GPa~\cite{16},
and experimental measurement 144~GPa~\cite{16}, but is larger than
the other theoretical value 62~GPa~\cite{23}. In general, the shear
modulus of cubic materials is slightly lower than the value of
$c_{44}$~\cite{16}, whereas our calculation indicates the opposite
case.

There have been proposals that the shear modulus may be a better
index of hardness~\cite{30}. Our calculated values of \textit{G},
225.98~GPa by LDA and 198.16~GPa by GGA, are much greater than the
theoretical values of 133~GPa~\cite{2} and 62~GPa~\cite{23}. Even
so, the current results are still incomparable with those of diamond
and \textit{c}-BN, e.g., the recently calculated values for diamond are
550~GPa~\cite{31}, 545~GPa~\cite{32}, 518~GPa~\cite{32}, and 403~GPa
for \textit{c}-BN~\cite{31}. However, the current results are
comparable to those of OsO$_2$~\cite{32} with values  of 250~GPa
(LDA) and 223~GPa (GGA) (\cite{15} suggests OsO$_2$ as a better
candidate for a hard material since their calculations confirmed that
the bulk modulus of OsO$_2$, 452~GPa, is only smaller than that of
diamond). Overall, it is reasonable to suggest that the fluorite
RuO$_2$ is a potential ultra-incompressible material, consistent
with the suggest superhard material from~\cite{2}.

In order to better understand the pressure responses of the mechanical
behavior, we have studied the bulk modulus \textit{B}, shear modulus
\textit{G}, Young's modulus \textit{Y}, and Lam\'{e}'s coefficients
$\lambda$ by increasing the pressure to 150~GPa. The Young's modulus
\textit{Y} and Lam\'{e}'s coefficients $\lambda$ are also essential for
understanding the macroscopic mechanical properties of solids and
for designing hard materials. Figure~\ref{fig:three} shows the most
significant pressure dependence of \textit{B} and the least significant
pressure dependence of \textit{G}. In contrast to \textit{B}, the
Lam\'{e}'s coefficients $\lambda$ increase slowly with pressure.
Compared to Lam\'{e}'s coefficients $\lambda$, the Young's modulus
\textit{Y} behaves much slower with the increase of pressure. At zero
pressure, the relative magnitude of the four mechanical parameters
in descending order is: $Y>B>G>\lambda$. However, the Lam\'{e}'s
coefficient $\lambda$ is larger than \textit{G} above nearly 30~GPa,
and the \textit{B} is larger than \textit{Y} above nearly 150~GPa. The high
\textit{Y} and \textit{B}, particularly at high pressures, also suggest that
fluorite RuO$_2$ is a potential ultrahard material.

The value of the Poisson's ratio for covalent materials is small
($\sigma=0.1$), whereas for ionic materials, a typical value of
$\sigma$ is 0.25~\cite{33}. In our cases, the value of $\sigma$ for
RuO$_2$ varies from about 0.2124 to 0.3227, as shown in table~\ref{tab:table2}, indicating a higher ionic and weaker covalent
contribution to intra-atomic bonding. Besides, the typical relation
between bulk and shear modulus is, respectively, $G\approx1.1B$ and
$G\approx0.6B$ for covalent and ionic materials. In our cases, the
calculated values of $G/B$ are in the range of 0.7396 at 0~GPa to
0.3656 at 150~GPa, indicating that the ionic bonding is dominant for
fluorite RuO$_2$. To evaluate the material ductility or brittleness,
Pugh et al. introduced the $B/G$ ratio~\cite{34}: the material
is brittle if the ratio is less than the critical value 1.75.
Therefore, fluorite RuO$_2$ is brittle under ambient conditions
since the $B/G$ is only 1.35. However, the brittleness decreases (or
ductility increases) when the pressure is increased and the $B/G$
ratio rises to 2.4 when the pressure is up to 120~GPa.

\begin{figure}[ht]
\centerline{\includegraphics[width=8cm]{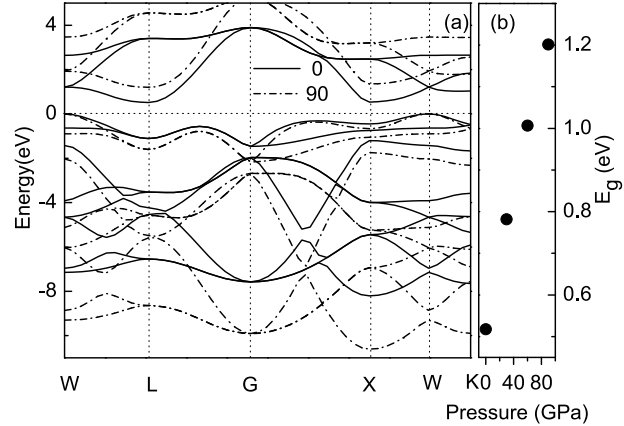}}
\caption{Energy band structure (by GGA method) along the high
symmetry points in the Brillouin zone at the pressures of 0, 90~GPa
is shown in (a) and the band gap $E_\mathrm{g}$ as a function with the
applied pressures is shown in (b). \label{fig:four}}
\end{figure}

\begin{figure}[ht]
\centerline{\includegraphics[width=8cm]{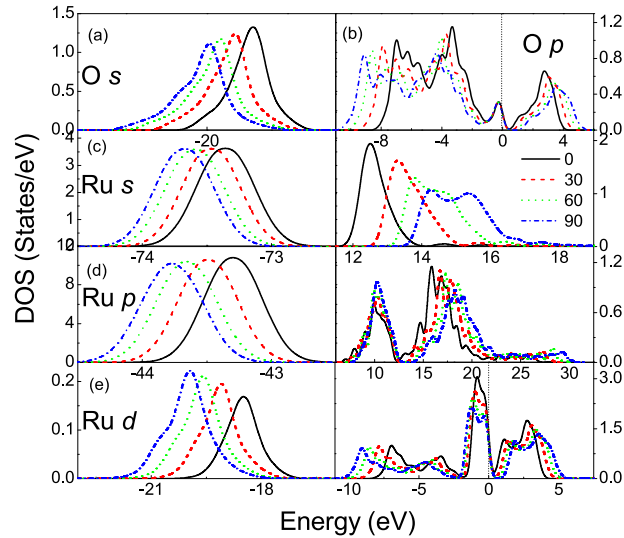}}
\caption{(Color online) Partial density of states (by GGA method) of O and Ru
states under $P=0$, 30, 60, 90~GPa. \label{fig:five}}
\end{figure}

\begin{figure}[ht]
\centerline{\includegraphics[width=8cm]{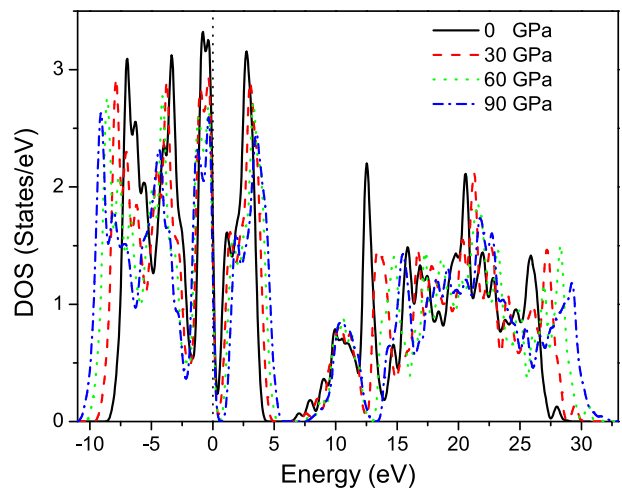}}
\caption{(Color online) Total density of states (by GGA method) under $P=0$, 30, 60,
90~GPa near the Fermi~level. \label{fig:six}}
\end{figure}

The dependences of the Debye temperature $\Theta$, heat
capacity $C_V$, Gr\"{u}neisen parameter $\gamma$, thermal
expansion coefficient $\alpha$, and Poisson's ratio $\sigma$ on
the pressure are calculated. As shown in table~\ref{tab:table2},
when the temperature keeps constant ($T=0$ K), $\Theta$ and
$\sigma$ increase with increasing the pressure, whereas $C_V$,
$\gamma$, and $\alpha$ decrease. The five thermodynamic
parameters show different pressure dependences within the range of
$0\div120$~GPa. It is obvious that the thermal expansion coefficient
$\alpha$ declines most significantly, corresponding to an 85\%
compression. The heat capacity $C_V$ and Gr\"{u}neisen parameter
$\gamma$, however, correspond to smaller compressions with 60\%
and 25\%, respectively. The other two parameters, Debye temperature
$\Theta$ and Poisson's ratio $\sigma$, increase with the
pressure with 32\% and 52\%, respectively. Moreover, all the five
parameters have shown decreased dependences with increasing the
pressure, indicating anharmonicity of the vibration. Therefore, it
is necessary to further investigate the electronic energy band
structure and density of states (DOS) to better understand the
physical properties. Accordingly, we have made a systematic
investigation of the fluorite RuO$_2$ at different pressures (0, 30,
60, 90~GPa) under 0 K, as shown in figures~\ref{fig:four},
~\ref{fig:five}, and~\ref{fig:six}.

Figure~\ref{fig:four} presents the pressure-induced energy level shift
towards higher and lower regions. We can see that the applied
pressure has a larger effect on the energy levels far away from the
Fermi level than those in the vicinity of the Fermi level,
indicating a stronger effect on the core level than on the valence
level. From the energy band structure, we find that the top of the
valence band occurs at \textit{W} point and the bottom of the
conduction band occurs at \textit{L} point (slightly lower than \textit{X}
point by a value of $0.02$~eV), implying that there exists
an indirect gap with width of $0.5175$~eV in  fluorite RuO$_2$. The calculated band gap at
zero pressure is consistent with the other theoretical value ($0.5$~eV)
\cite{2}, but is much smaller than those of diamond ($4.15$~eV) and
\textit{c}-BN ($4.49$~eV)~\cite{35}. Moreover, it is found that the band
gap increases almost linearly with the pressure.

In figures~\ref{fig:five} and~\ref{fig:six}, we plot the calculated
DOS by GGA, where the Fermi energy is taken to be zero. From
figure~\ref{fig:five} (a), the O 2\textit{s} band center is at $-18.5$~eV,
which is consistent with those in rutile~\cite{10,14} and
CaCl$_2$-type~\cite{14} RuO$_2$. The valence band width is about 4
eV, which is much larger than those in rutile~\cite{14} (2.5
eV) and CaCl$_2$-type~\cite{14} RuO$_2$ (1.4~eV). In
figure~\ref{fig:five} (b), the calculated valence band width of O
2\textit{p} is about 8~eV, which is slightly larger than that in rutile
RuO$_2$ of 5.9~eV (using an extended linear augmented plane wave
potential) and 6.8~eV (using linear-muffin-tin-orbital potential)
\cite{10}, but the present calculation is consistent with that in
rutile~\cite{14} and CaCl$_2$-type~\cite{14} RuO$_2$ with the same
values of about 8.1~eV. Furthermore, the calculated conduction band
width is 4~eV, which is far smaller than that of valence conduction.

The Ru \textit{s} semi-core band, centered at $-73.3$~eV, displays larger
relative intensity (3.6) and smaller width (1.2~eV) with respect to
those in the conduction band with smaller relative intensity (1.9) and
larger width (2.2~eV). The other Ru \textit{s} electrons distribute
mainly in the ranges of $-21~\text{eV}\div -18~\text{eV}$, $-8.6~\text{eV}\div -1.6~\text{eV}$,
$-1.5~\text{eV}\div0.5~\text{eV}$, have formed very weak peak with intensities
less than 0.1, and thus could be ignored as compared with the peak
far away from the Fermi level. In figure~\ref{fig:five} (d), the Ru
\textit{p} state locates at $-43.5$~eV in the valence band with width of
1.3~eV and distributes in the energy range of 6.3$\div$27.5~eV in
the conduction band (corresponding to two sharp peaks centered at
around 10~eV and 16~eV, respectively). The relative intensity of Ru
$p$ state in the valence band is far greater than that in
conduction band, and the DOS distributed in the energy range
$-20 \div 5$~eV is ignored due to their subtle relative
intensities (below 0.12). The Ru 4\textit{d} state, shown in
figure~\ref{fig:five} (e), is distributed at $-18.8$~eV and around the
Fermi level with widths of 2.3~eV and 12.0~eV, respectively.
Moreover, the relative intensity of the inner valence band is
considerably smaller than that in the outer valence band and in the
conduction band.

The four sharp peaks of the total DOS within $-7.5 \div 5$~eV
originate from the strong hybridization between Ru 4\textit{d} and O
2\textit{p}, as seen in figure~\ref{fig:six}. The complete overlap of the
Ru 4\textit{d} and O 2\textit{p} states from $-8$~eV to 4~eV indicates a
strong covalent interaction between them, whereas the nonzero DOS
value at Fermi level indicates a moderate metallic feature in its
bonding state. Although there is a large hybridization between Ru
4\textit{d} and O 2\textit{p} states, the charge transfer from Ru to O is
possible in the present case. By analyzing the Mulliken population
results, it is found that the charge transfer from Ru to O is as
numerous  as about 1.01 electrons. Therefore, the bonding behavior
between Ru-O has ionic contributions owing to charge transfer. In a
word, the bonding behavior between the Ru-O is a combination of
covalent, metallic and ionic contributions.

To emphasize the pressure dependence of the DOS, we have
investigated the DOS under different pressures (30, 60, 90~GPa), as
shown in figures~\ref{fig:five} and~\ref{fig:six}. It is clearly seen
that the applied pressure causes the energy levels shifting towards
both sides of the Fermi level and thus the energy band is broadened.
Under a higher pressure, the energy level shift is decreased, implying
the strong repulsion among the core electrons. Meanwhile, in
general, the relative intensity decreases with increasing the
pressure. Accordingly, the relative shift in the lower energy space
is always larger than those in the higher energy space, implying
different bonding strength. The change of the DOS can be attributed
to the charge transfer during lattice distortion. With the increase
of pressure, a higher overlap of the wavefunction results in a
stronger delocalization of electrons. Electrons transfer from
the majority to minority spin band and form broader bands. The
center changes and electrons become more localized when lattice
distortion changes from negative to positive. The majority and
minority bands move with respect to the Fermi level, which
 affects both the spin polarization ratio and magnetic
properties. The current investigations reveal that the relative
intensity of O \textit{s} and \textit{p} states decrease with the pressure
both in the valence and conduction bands. However, the relative
intensity of Ru \textit{s} state keeps almost unchanged in the
semi-core band and the main peak in the conduction band has been
split into two peaks with the pressure. By analyzing Ru \textit{p}
state, we find that the relative intensity decreases slightly with
the pressure in the whole valence band and in the higher-energy
range from 12.5~eV to 25~eV, but increases with the pressure within
7.5~eV$\div$12.5~eV. Interestingly, there is observed an increase of the relative
intensity of Ru \textit{d} state with the pressure in the deeper-lying
valence band, whereas the relative intensity of the
other Ru \textit{d} state decreases with the pressure, presenting
opposite variation tendencies. The different intensity variation
trends have unambiguously demonstrated that the applied pressure has
induced various charge transfer tendencies.

\section{Conclusions}

The current investigations revealed that the fluorite RuO$_2$ is a
potential ultrahard material. The elastic stability criteria show
that the fluorite RuO$_2$ is elastic stable up to 120~GPa. The
calculated Poisson's ratio and Debye temperature increase
monotonously with the pressure. However, the heat capacity, Gr\"{u}neisen
parameter, and thermal expansion coefficient decrease with the
pressure. An analysis based on Poisson's ratio, $G/B$, and DOS
reveals that the bonding nature in RuO$_2$ is a combination of
ionic, covalent and metallic contributions, which contributes to the
hardness and fundamental properties. The energy band investigations
found an energy gap between the top of the valence band and the
bottom of the conduction band, and the gap seems to increase
monotonously with the pressure. Moreover, the different intensity
variation trends of DOS have unambiguously demonstrated that the
applied pressure has caused various charge transfer tendencies.

\section*{Acknowledgements}
We are thankful for financial support from the National Natural
Science Foundation of China \linebreak (Grant Nos: 10974139, 10964002,
11104247) and the Provincial Natural Science Foundation of Guizhou
(Grant Nos: [2009]2066 and TZJF--2008--42), Hainan (Grant No: 110001),
Chong Qing (Grant \linebreak No:~CSTCcstc2011jja90002) and Zhejiang (Grant No:~Y201121807).

\newpage

\ukrainianpart

\title{Пружні та електронні властивості флюориту RuO$_2$ \\ з перших принципів }
\author{З.Дж. Янг\refaddr{ad1}, А.M. Гуо\refaddr{ad2},
Й.Д. Гуо\refaddr{ad3}, Дж. Лі\refaddr{ad4}, З. Ванг\refaddr{ad4}, К.
Ліу\refaddr{ad5}, Р.Ф. Лінгха\refaddr{ad6}, X.Д. Янг\refaddr{ad7}
}

\addresses{\addr{ad1}Природничий факультет, Технологічний університет Чжецзяну, 310023 Ханчжоу, КНР
\addr{ad2} Фізичний факультет, Каліфорнійський державний університет,
Нортрідж,   Каліфорнія 91330--8268, США
\addr{ad3}Фізичний факультет, Педагогічний університет Нейцзяню,  Нейцзян 641112, КНР
\addr{ad4}Коледж матеріалознавства і хімічної інженерії, Xайнанська
центральна регіональна дослідна лабораторія з питань утилізації Si-Zr-Ti, Xайнанський університет,  Хайкоу 570228, КНР
\addr{ad5}Фізичний факультет, Чунцинський технологічний університет,
 Чунцин 400050, КНР
\addr{ad6} Фізичний факультет, Педагогічний університет Гуйчжоу,  Гуйян 550001, КНР
\addr{ad7}Інститут атомної і молекулярної фізики, Сичуанський
університет,  Ченду 610065, КНР
}

\makeukrtitle

\begin{abstract}
\tolerance=3000%
Пружні, термодинамічні та електричні властивості флюориту RuO$_2$
при високому тиску досліджуються за допомогою теорії функціоналу
густини з плоскохвильовим псевдопотенціалом. Оптимізовані параметри
гратки, пружні сталі, об'ємний модуль і модуль зсуву  узгоджуються з
іншими теоретичними значеннями. Фазовий перехід з модифікованого
флюориту до флюориту є при 88~GPa (наближення локальної
густини, LDA), чи при 115.5~GPa (узагальнене градієнтне наближення,
GGA). Також досліджено модуль Юнга і коефіцієнти Ламе при високих
тисках. Структура є стабільною для тисків до  120~GPa, якщо
обчислювати пружні сталі. Крім того, досліджено термодинамічні
властивості, включаючи температуру Дебая, теплоємність, коефіцієнт
теплового розширення, параметр Грюнайзена і коефіцієнт Пуассона.
В електронній структурі флюориту  RuO$_2$ знайдено малу зонну щілину
і ширина зони зростає із тиском. Також, представлені механічні та
електронні властивості демонструють, що природа зв'язування є
комбінацією ковалентного, іонного і металічного вкладів.
\keywords перші принципи, електронна структура, пружність, термодинамічність


\end{abstract}

\end{document}